\documentclass[graybox]{svmult}

\usepackage{mathptmx}       
\usepackage{helvet}         
\usepackage{courier}        
\usepackage{type1cm}        
\usepackage{makeidx}         
\usepackage{graphicx}        
\usepackage{multicol}        
\usepackage[bottom]{footmisc}

\usepackage[sort]{natbib}
\usepackage{gal_post_risquez_1_aas_macros}

\usepackage{url}

\makeindex             

\begin{document}

\title*{High Energy Sources Monitored with OMC}
\author{D.~Risquez \and A.~Domingo \and M.D.~Caballero-Garc\'ia \and J.~Alfonso-Garz\'on \and J.M.~Mas-Hesse}
\authorrunning{D.~Risquez et al.}
\institute{All authors are members of Centro de Astrobiolog\'ia -- LAEFF (CSIC--INTA), Apartado 78, E-28691 Villanueva de la Ca\~nada, Madrid, Spain, \\ \url{omc-support@laeff.inta.es}}


%
%
\maketitle


\abstract{The Optical Monitoring Camera on-board INTEGRAL (OMC) provides Johnson V band photometry of any potentially variable source within its field of view. Taking advantage of the INTEGRAL capabilities allowing the simultaneous observation of different kind of objects in the optical, X and gamma rays bands, we have performed a study of the optical counterparts of different high-energy sources. Up to now, OMC has detected the optical counterpart for more than 100 sources from the High Energy Catalog \citep{Ebisawa2003}. The photometrically calibrated light curves produced by OMC can be accessed through our web portal at: \hbox{\url{http://sdc.laeff.inta.es/omc}}}

\section{Introduction}

The INTEGRAL Optical Monitoring Camera, OMC \citep{Mas-Hesse2003a}, observes the optical emission from the prime targets of the gamma ray instruments on-board the ESA mission INTEGRAL: 
SPI (gamma ray spectrometer) and IBIS (gamma ray imager), with the support of the JEM-X monitor in the X-ray domain. OMC has the same field of view (FOV) as the fully coded FOV of JEM-X, and it is co-aligned with the central part of the larger fields of view of IBIS and SPI. This capability provides invaluable diagnostic information on the nature and the physics of the sources over a broad wavelength range.

The OMC is based on a refractive optics with a Johnson V~filter pass-band (centred at 550~nm). It has an aperture of 50~mm focused onto a large format CCD ($1024 \times 2048$~pixels) working in a frame transfer mode ($1024 \times 1024$~pixels imaging area). Its field of view is $5^\circ \times 5^\circ$ and the image scale is $17.5$~arcsec/pixel.

%
%

All the data processed by OMC is open to the scientific community. The OMC team has developed a scientific archive \citep{Gutierrez2004}, containing the data generated by the OMC and an access system capable of performing complex searches, complementary to the INTEGRAL Archive hosted at ISDC. It is reachable at: \url{http://sdc.laeff.inta.es/omc}

\section{Some Monitored Sources}

%

\subsection{Cyg~X-1}

This source is a binary system constituted by an O9.7Iab star and a compact object, potentially a black hole. The optical counterpart is known as HD~226868 \citep{Bolton1972,Webster1972}.

It is a HMXB (High Mass X-ray Binary), and it is well known that it displays a double-peaked ellipsoidal modulation as expected from a tidally and rotationally distorted star. This tidal distortion is induced by the presence of the massive, though not visible, companion.


A significant degree of variability over more than 4~years is evident in the OMC light curve.


The average amplitude of the optical variations of $0.06$~mag and the difference between maxima of $0.015$~mag confirm earlier studies by \cite{Bruevich1978}. Small differences and changes probably reflect variations in the activity of the X-ray source. The difference between maxima could be attributed to a non-uniform distribution of the surface brightness of the optical star.


\subsection{GX~301-02}

This is another HMXB, with a Be super-giant companion \citep{White1976} and an orbital period of $41.498$~days \citep{Bildsten1997}.

\begin{figure}[t]
\includegraphics[width=0.42\textwidth]{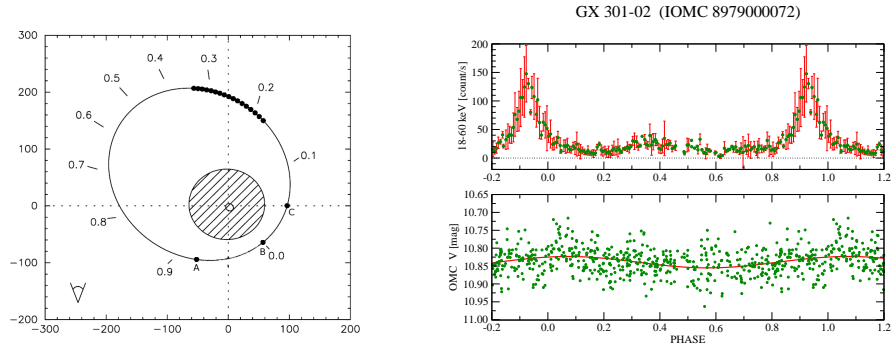}
\hfill
\includegraphics[width=0.499\textwidth]{gal_post_risquez_1_fig2.eps}
\caption{GX~301-02. On the left it is displayed the orbital sketch. It has been modified from \cite{Kaper2006}. On the right, the gamma (top) and optical (bottom) light curve. The line in the optical plot is a simple sinusoidal fit to the data.}
\label{1:fig:GX301-02}
\end{figure}

In the Fig.~\ref{1:fig:GX301-02} we use INTEGRAL data to justify the physical model of the system. The peak of high energy emission takes place in point A. B is the periastron. The peak of optical emission occurs around C, when the companion star faces the largest cross-section on the line of sight.

The gamma ray light curve shows two maxima. They happen when the compact object, in its orbital movement, goes through the disk outflow around the super-giant companion. Disk density is higher near the companion star, and due to that the primary maximum is brighter than the secondary one. The optical curve shows a weak variability, but it is exactly as expected by the physical model.


\subsection{Her~X-1}

This source is also known as HZ~Her. It is an LMXB composed of a neutron star and a companion star of $2.3$~solar masses circling each other every $1.7$~days in an almost circular orbit. X-ray eclipses have been detected when the neutron star is hidden by its companion.

Optical variations are due to the tidal distortion of the companion star, and also to the intense X-ray heating of the illuminated face of the companion produced by the neutron star \citep{Reynolds1997}.

Figure~\ref{1:fig:HerX1andIGR_J17497-2821}, left side, displays the optical light curve. The insets show an artist's impression of the system and a sketch of the situation at different orbital phases (not to scale and omitting the accretion disk).

\section{Galactic Bulge Monitoring}

The Galactic Bulge is a region rich in bright variable high-energy X-ray and gamma-ray sources. The OMC team is involved in the Galactic Bulge Monitoring \citep{Kuulkers2007}. This project consists in some periodic and systematic observations of this region of the Galaxy with several instruments on-board INTEGRAL: IBIS/ISGRI, JEM-X and OMC. The main goal of this monitoring is to obtain simultaneous light curves in different energy ranges. Results (at the moment just for high energies) are publicly available at:\\ \hbox{\url{http://isdc.unige.ch/Science/BULGE}}

OMC monitors all sources detected by IBIS/ISGRI in the Galactic Bulge which are in its FOV. The full analysis is working in an automatic way, producing the results (light curves) in a short period of time. In this way, if a flare occurs, it could be detected immediately with OMC.

\begin{figure}[t]
\includegraphics[width=0.495\textwidth]{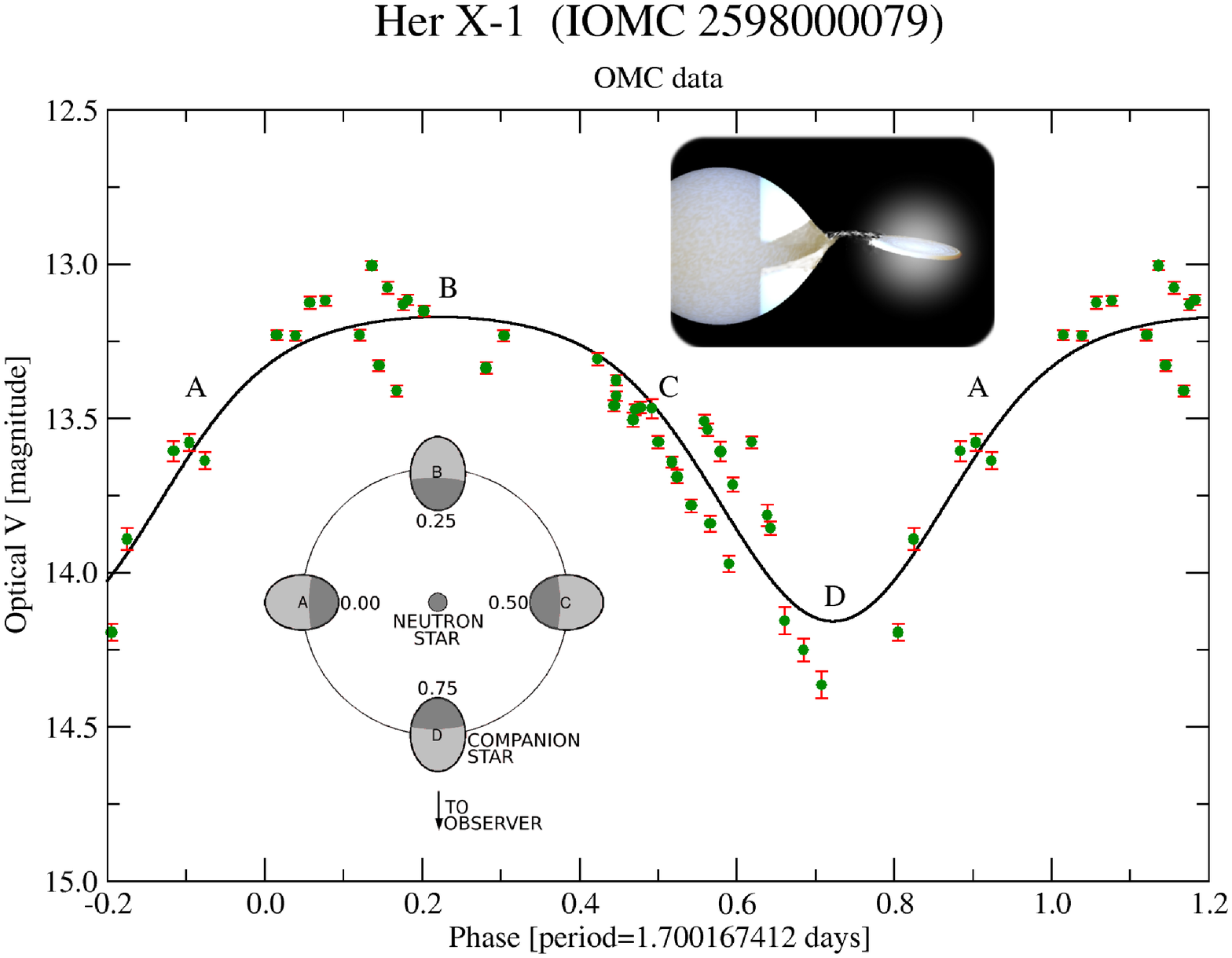}
\includegraphics[width=0.495\textwidth]{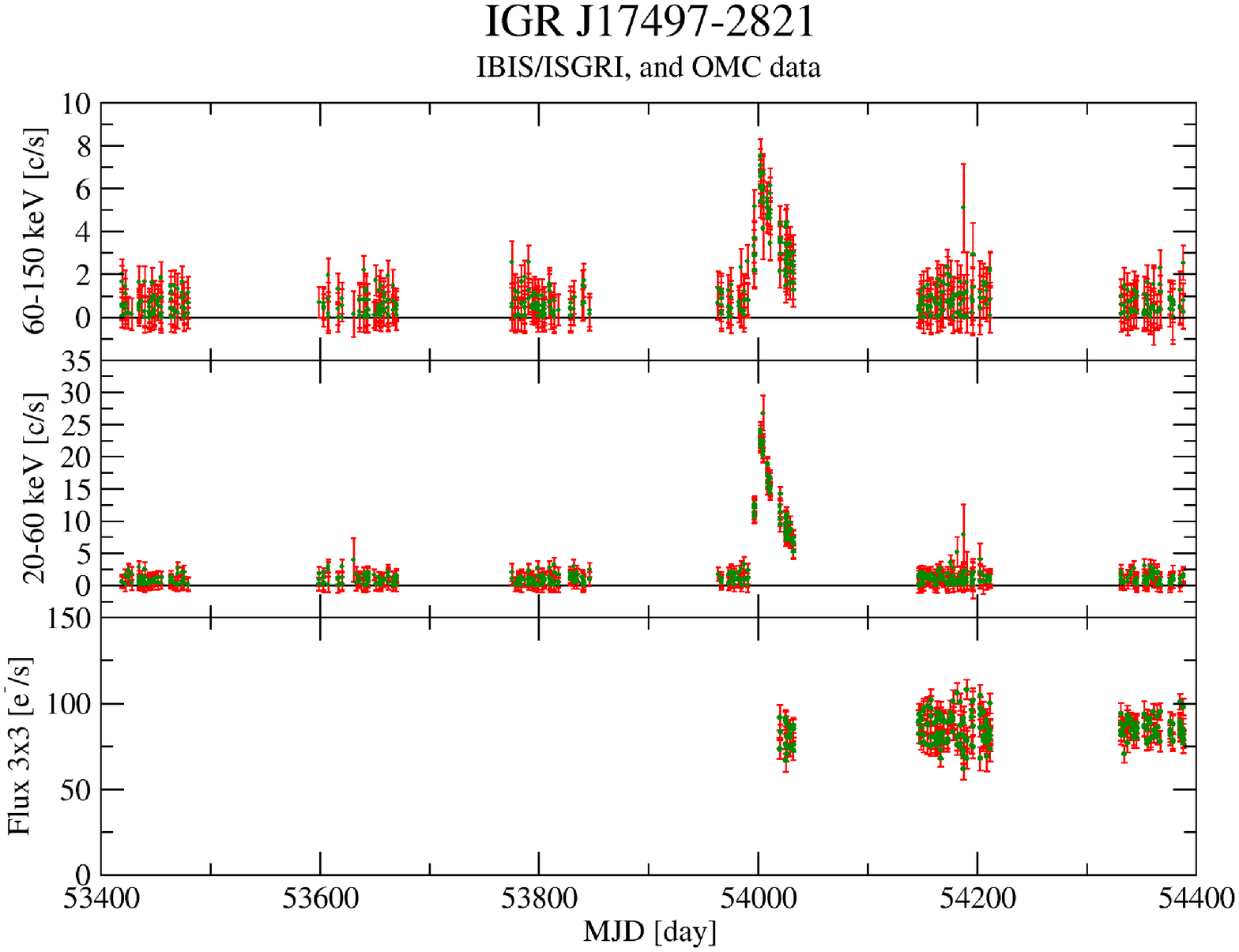}
\caption{On the left: Her~H-1 light curve and orbital sketch. On the right: IGR~J17497-2821 light curves obtained during the Galactic Bulge Motoring. Top and middle light curves are in the gamma ray energy range. The bottom light curve has been obtained with OMC.}
\label{1:fig:HerX1andIGR_J17497-2821}
\end{figure}

As an example we show IGR~J17497-2821 (Fig.~\ref{1:fig:HerX1andIGR_J17497-2821}, right side). This source was discovered by \cite{Soldi2006}. Unfortunately, when the flare was detected by the high energy instruments on-board INTEGRAL, OMC was not observing this region.

\begin{acknowledgement}
The activities related to INTEGRAL-OMC are being funded since 1993 by the Spanish National Space Programme (MEC/MICINN).
\end{acknowledgement}


\bibliographystyle{gal_post_risquez_1_aa}
\bibliography{gal_post_risquez_1_cites}


\end{document}